\documentclass[aps, prb, preprint, superscriptaddress]{revtex4}
\usepackage[english]{babel}
\usepackage{amsmath,amsthm}
\usepackage{amsfonts}
\usepackage{xspace}
\usepackage{bm}
\usepackage[squaren, thinqspace]{SIunits}
\usepackage{booktabs}
\usepackage{graphicx}

\begin{document}
\newcommand{\w}{\ensuremath{W\xspace}}
\newcommand{\Hres}{\ensuremath{H_{n}^{\text{res}}}\xspace}
\newcommand{\Hex}{\ensuremath{H_{n}^{\text{ex}}}\xspace}
\newcommand{\Ms}{\ensuremath{M_{\text{s}}}\xspace}
\newcommand{\mx}{\ensuremath{m^{x}_{n}(H_{0},I;z)}\xspace}
\newcommand{\A}{\ensuremath{A_{n}}\xspace}
\newcommand{\alphae}{\ensuremath{\alpha^{\text{eddy}}}\xspace}
\newcommand{\alphar}{\ensuremath{\alpha_{n}^{\text{rad}}}\xspace}
\newcommand{\alphaint}{\ensuremath{\alpha^{\text{int}}}\xspace}
\newcommand{\Ln}{\ensuremath{\widetilde{L}_{n}}\xspace}
\newcommand{\hx}{\ensuremath{h^{x}(I;x,z)}\xspace}
\newcommand{\intx}{\ensuremath{\int^{\infty}_{-\infty}dx}\xspace}
\newcommand{\intz}{\ensuremath{\int^{\delta+d}_{d}dz}\xspace}
\newcommand{\Phin}{\ensuremath{\Phi_{n}\left(H_{0},I\right)}\xspace}
\newcommand{\Sto}{\ensuremath{\Delta S^{21}_{n}}\xspace}
\newcommand{\Vin}{\ensuremath{V_{\text{in}}}\xspace}
\newcommand{\muB}{\ensuremath{\mu_{\text{B}}}\xspace}

\title{Radiative damping in wave guide based FMR measured via analysis of perpendicular standing spin waves in sputtered Permalloy films}

\author{Martin A. W. Schoen}
\affiliation{Institute of Experimental and Applied Physics, University of Regensburg, 93053 Regensburg, Germany}
\author{Justin M. Shaw}
\affiliation{Electromagnetics Division, National Institute of Standards and Technology, Boulder, CO, 80305}
\author{Hans T. Nembach}
\affiliation{Electromagnetics Division, National Institute of Standards and Technology, Boulder, CO, 80305}
\author{Mathias Weiler}
\affiliation{Walther-Meißner-Institut, Bayerische Akademie der Wissenschaften, D-85748 Garching, Germany}
\author{Thomas J. Silva}
\affiliation{Electromagnetics Division, National Institute of Standards and Technology, Boulder, CO, 80305}
\email{martin1.schoen@physik.uni-regensburg.de}
\date{\today}%

\begin{abstract}

The damping $\alpha$ of the spinwave resonances in \unit{75}{\nano\meter}, \unit{120}{\nano\meter}, and \unit{200}{\nano\meter} -thick Permalloy films is measured via vector-network-analyzer ferromagnetic-resonance (VNA-FMR) in the out-of-plane geometry. Inductive coupling between the sample and the waveguide leads to an additional radiative damping term. The radiative contribution to the over-all damping is determined by measuring perpendicular standing spin waves (PSSWs) in the Permalloy films, and the results are compared to a simple analytical model. The damping of the PSSWs can be fully explained by three contributions to the damping: The intrinsic damping, the eddy-current damping, and the radiative damping. No other contributions were observed. Furthermore, a method to determine the radiative damping in FMR measurements with a single resonance is suggested.
\vfill
\center{Contribution of NIST, not subject to copyright}
\end{abstract}

\maketitle

\section{Introduction}
Excited magnetic moments relax towards their equilibrium orientation due to damping. Several physical mechanism can cause damping. Many mechanisms, like eddy current damping~\cite{Lock1966} in conducting ferromagnets, were already identified in the 1950s. More recently, enhanced damping due to spin pumping~\cite{Tserkovnyak2002} from a ferromagnet into an adjacent metallic layer was identified, and remains a topic of ongoing investigation\cite{Y.Tserkovnyak2005,Y.Tserkovnyak2002,Boone2013,Shaw2012}. Furthermore, wavenumber-dependent contributions to the damping caused by intralayer spin pumping have been theoretically predicted~\cite{Tserkovnyak2009a,V.Baryakhtar1986} and currently are the subject of experimental investigation~\cite{Li2014,Nembach2013}. 
Another damping process, referred to as radiative damping\cite{Sanders1974,Wende1976}, has been known to exist since the 1970s and is purely due to inductive coupling between the sample and the waveguide in ferromagnetic resonance (FMR) experiments. More recently, this phenomenon has been further investigated in the context of strong magnon-photon coupling experiments, with possible applications in quantum information processing\cite{Bhoi2014}. In these quantum-coherent experiments, radiative damping was identified as a manifestation of non-resonant magnon-photon coupling, and it was determined that such coupling\cite{Bai2015} is indeed a source of extrinsic line width in cavity-based FMR studies. 
In a radiative damping process, the time-varying magnetic flux associated with the dynamic magnetization generates microwave-frequency currents in the proximate conductor of a microwave waveguide that carries the resultant power away from the sample. This process is similar to that exploited in eddy-current brakes and can be seen as a non-local counterpart to the eddy-current damping in conductive ferromagnets. To determine the magnitude of radiative damping in magnetic thin films, we used broadband vector-network-analyzer ferromagnetic resonance (VNA-FMR) to measure damping in $\text{Ni}_{0.8}\text{Fe}_{0.2}$ Permalloy (Py) films with thicknesses $\delta$ varying between \unit{70}{\nano\meter} and \unit{200}{\nano\meter}. By use of the geometry sketched in Fig.~\ref{fig:rad}(a), we determine the total damping for each mode $\alpha_{n}$ as a sum of intrinsic damping $\alpha^{\text{int}}$, eddy-current damping \alphae and radiative damping \alphar.
%
%
We then perform a quantitative analysis of the PSSW resonance fields, amplitudes and damping to extract the different contributions to $\alpha_{n}$. We find that eddy current damping is only significant for the lowest order mode, the radiative damping strongly affects the first five modes, and no additional contributions to the damping are detectable for spin waves up to $k=\unit{1.75\times10^{6}}{\centi\meter^{-1}}$. This last finding is in contrast with reports of exchange mediated damping in both nanostructures\cite{Nembach2013} and thin films\cite{Li2014}.

\section{damping models}

\begin{figure}[!htp]
\begin{center}
		\includegraphics [width=0.7\textwidth] {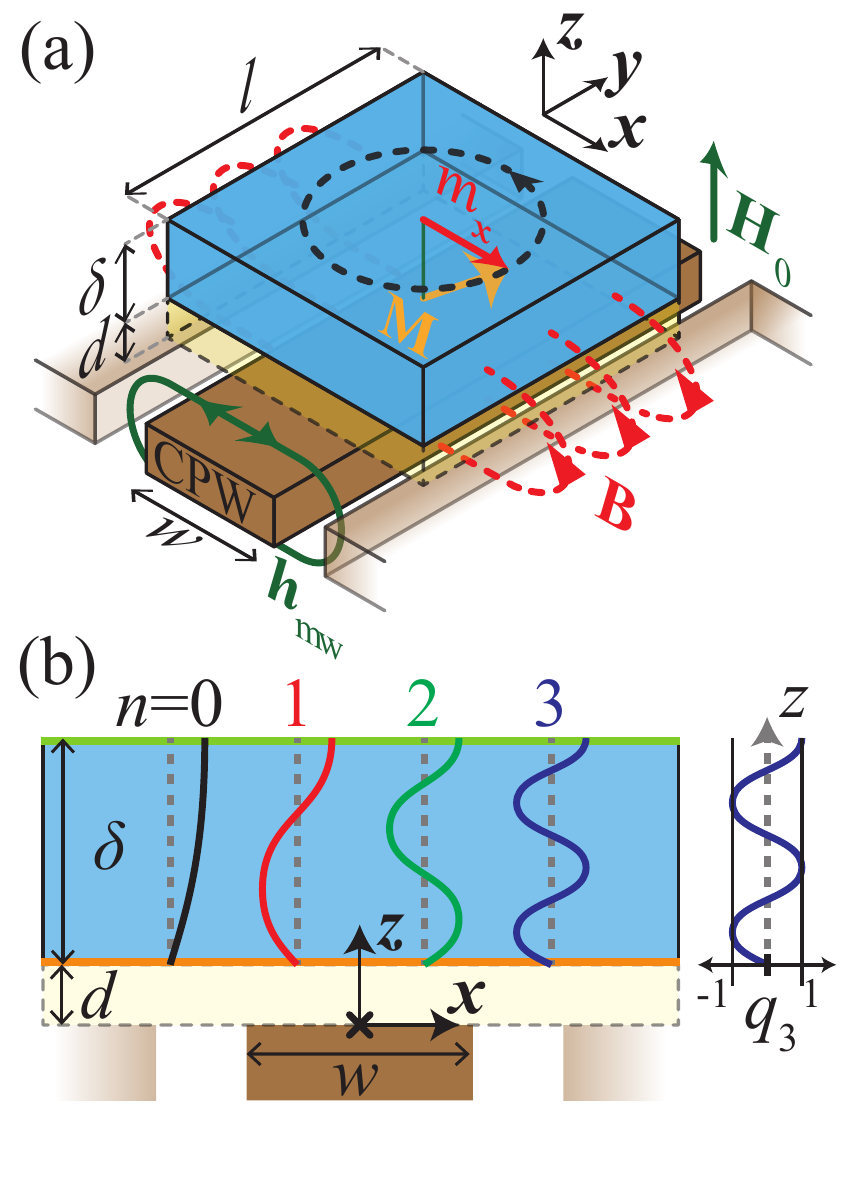}
		\caption{Schematic of the radiative damping process. (a) $\bm{M}$ is the dynamic magnetization, $\bm{H}_{0}$ the applied external field and $\bm{B}$ is the magnetic inductance due to the $x$-component $m_x$ of the dynamic magnetization. $\w$ is the width of the center conductor, $l$ the length of the sample on the waveguide, $\delta$ the thickness of the sample and $d$ the spacing between sample and wave guide. (b) Simplified depiction of the PSSW eigenfunctions $q_{n}$ for mode numbers n=~0,~1,~2,~3. We exemplarily used boundary conditions that are completely pinned on one side and completely un-pinned on the other side. The origin of the coordinate system is indicated.}
\label{fig:rad}
\end{center}
\end{figure}

According to Faraday's Law, the time-varying flux of a precessing magnetic moment generates an ac voltage in any conducting material that passes through the flux. As shown in Fig.~\ref{fig:rad}, spin wave precession in a conducting ferromagnet on top of a coplanar waveguide (CPW) induces ac currents both in the ferromagnet and the CPW. The dissipation of these eddy currents in the sample, and the flow of energy away in the CPW give rise to two contributions to magnetic damping. Historically, the damping caused by eddy currents in the ferromagnet $\alphae$ is called eddy current damping, while the damping caused by the eddy currents in the waveguide is called radiative damping $\alphar$.

Eddy current damping has been recognized since the 1950s \cite{Lock1966,Scheck2006}. For the lowest order mode in FMR,
\begin{equation}
\alphae = \frac{C}{16}\frac{\gamma \mu_{0}^{2}M_{\text{s}}\delta^{2}}{\rho},
\label{eddy}
\end{equation} 
with the resistivity $\rho$, saturation magnetization $M_{s}$, the vacuum permeability $\mu_{0}$, the gyro-magnetic ratio $\gamma$, and the sample thickness $\delta$ (see derivation in Appendix Sec. B).  We introduce a correction factor $C$ to account for details of the eddy current spatial profile. As shown in a later section, $\alphae$ for all higher order PSSW modes investigated in this study is much smaller than that of the lowest order mode.

We now turn to the radiative damping $\alphar$. 
We consider the experimental geometry sketched in Fig.~\ref{fig:rad}~(a). A ferromagnetic sample with thickness $\delta$ and length $l$ is placed on top of the center conductor of a coplanar waveguide with width \w. The sample dimension along $\bm{x}$ is much larger than \w. The sample and CPW are separated by a gap of height $d$. An external dc magnetic field $\bm{H}_0$ is applied perpendicular to the sample plane, and the spin wave resonances (SWR) are driven by microwaves in the CPW at resonance frequency $f$. A fraction of the ac magnetic induction $\bm{B}$ due to the dynamic component of the magnetization \mx wraps around the center conductor. 

To derive a quantitative expression for \alphar, we start by calculating $\hx$, the $\bm{x}$ component of the driving field $\bm{h}_\mathrm{mw}$ that is generated by an excitation current $I$ in the center conductor. We assume $\hx$ is uniform along $\bm{y}$, but we allow for variation along $\bm{x}$ and $\bm{z}$. To estimate $\hx$ we use the Karlqvist equation~\cite{Mallinson1993} 
\begin{align}
\begin{split}
\hx= \frac{I}{2\pi \w}\left[\text{arctan}\left(\frac{x+\w/2}{z}\right)-\text{arctan}\left(\frac{x-\w/2}{z}\right)\right].
\label{kquist}
\end{split}
\end{align}
This microwave field can excite PSSWs in the sample. Schematic mode profiles for the fundamental mode ($n=0$) and the first three PSSW modes are shown in Fig.~\ref{fig:rad}(b), where we use unpinned boundary conditions at the top surface and pinned boundary conditions at the bottom surface. As shown in Fig.~\ref{fig:rad}(b), the mode profiles describe a z-dependence of the dynamic magnetization components $m^x$ and $m^y$. In the perpendicular geometry used here, $|m^x|=|m^y|$ everywhere, i.e. the precession is circular. In what follows, we will only discuss $m^x$, the dynamics of which are inductively detected in the measurement. For a PSSW with mode number $n$,  $\widetilde{m}^x_n(x,z)=q_n(z) \chi_n  \left\langle q_n(z) \hx \right\rangle$ where $\langle\rangle$ denotes spatial averaging in $x$ and $z$ directions, as defined in the Appendix, $\chi_n=\chi_n^{xx}$ is the diagonal component of the magnetic susceptibility of the $n$-th order mode, and $-1\leq q_n(z)\leq 1$ is the normalized mode profile (eigenmode), an example of which is sketched in Fig.~\ref{fig:rad}~(b) for $n=3$.
The mode inductance $L_{n}$ is given by $L_{n}=\chi_{n}\Ln$, where, as detailed in the Appendix, we define a normalized mode inductance $\Ln$ for the $n$th PSSW mode,
\begin{align}
\begin{split}
\Ln = \frac{\mu_{0}l}{I^2} \left\langle q_n(z) \hx \right\rangle^2 \w \delta\;.
\label{z}
\end{split}
\end{align}

\Ln, as explained in the appendix, no longer has any dependence on magnetic field or excitation frequency. In the simplest case of a uniform magnetization profile $q_{0}(z)=1$ (FMR-mode) and uniform excitation field $h^x(I;x,z)=h^x(I;0,0)=I/(2 \w)$, the normalized inductance is $\Ln = \mu_0 \delta l/(4 \w)$.

The $x$-component of the dynamic magnetization $m^{x}_{n}(x,z)$ produces a net flux $\Phi_{n}=\chi_{n}I\Ln$ that threads around CPW center conductor, leading to a power dissipation 
\begin{equation}
P_{n}= \frac{\omega^2 }{2Z_{0}}\left(\chi_{n}I\Ln\right)^{2},
\label{dissln}
\end{equation}
where $Z_0$ is the waveguide impedance (in our case $Z_0=\unit{50}{\ohm}$) and $\omega$ is the angular frequency of the magnetization precession.
With Eq.~\eqref{dissln}, the power dissipation rate $\left(\frac{1}{T_{1}}\right)_{n}=P_{n}/E_{n}$ can be calculated, where $E_{n}$ is the energy of the dynamic component of the magnetization derived in the Appendix. This power flow from the sample to the waveguide leads to the radiative damping contribution
\begin{align}
\begin{split}
\alpha^{\text{rad}}_{n}= \frac{1}{2\omega}\left(\frac{1}{T_{1}}\right)_{n}=\eta \gamma \mu_{0} M_{s} \frac{\Ln}{Z_{0}},\\ 
\label{rad}
\end{split}
\end{align}
where $\eta=\delta/\left(4\int^{\delta}_{0}dz\left|q_{n}(z)\right|^{2}\right)$ is a dimensionless parameter that accounts for the actual mode profile in the sample, see Appendix~Sec.~A. In the case of sinusoidal PSSWs, $\eta = 1/2$, and for a completely uniform mode profile, i.e. $q_{n}(z)=1$, $\eta = 1/4$. 
From Eq.~\eqref{rad}, it is evident that $\alphar$ is proportional to $\Ln$ for $n>0$. In the simplest case of uniform driving field $h^x=I/(2 \w)$, the radiative contribution is given by 
\begin{equation}
\alpha^{\text{rad}}_{0}=\frac{\eta\gamma\mu_{0}\Ms}{Z_{0}}\Ln\cong\frac{\eta\gamma\Ms\mu_{0}^{2}\delta l}{2Z_{0}\w}\;.
\label{simplealpha}
\end{equation}
Note that the radiative damping thus depends on the sample and waveguide dimensions, in particular linearly on the sample thickness. Unlike eddy-current damping, $\alphar$ is independent of the conductivity of the ferromagnet, hence this damping mechanism is also operative in ferromagnetic insulators. 

\section{Samples and method}

\begin{figure}[!htp]
\begin{center}
		\includegraphics [width=0.8\textwidth] {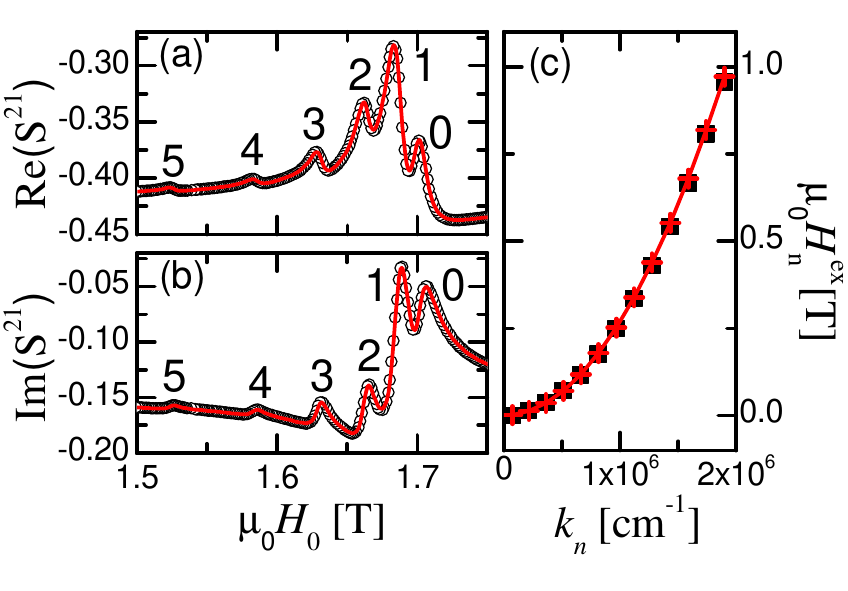}
		\caption{Measured $S_{21}$ transmission parameter (black circles) at 20 GHz and the multi-peak -susceptibility fit (red line) for the (a) real part and (b) imaginary part obtained with the Ta(3)-Py(200)-$\text{Si}_{3}\text{N}_{4}$(3) sample. The first 6 modes are shown. (c) The exchange field $H_{n}^{\text{ex}}$ (black squares) and exchange field fit, from Eqs.~\eqref{hex}~and~\eqref{soohoo} (red crosses) for all 13 detected modes plotted as a function of the fitted wave numbers $k_{n}$.}
		
		\label{spectrum}
\end{center}
\end{figure}

We deposit Ta(3)/Py($\delta$)/$\text{Si}_{3}\text{N}_{4}(3)$, Ta(3)/Py($\delta$)/Ta(5), Ta(3)/Py($\delta$) and Py($\delta$) layers on \unit{100}{\micro\meter} thick glass substrates by DC magnetron sputtering at a Ar pressure of \unit{0.7}{\pascal} ($\approx5\cdot10^{-3}\text{Torr}$) in a chamber with a base-pressure of less than \unit{5\cdot10^{-6}}{\pascal} ($\approx4\cdot10^{-8}\text{Torr}$); where $\delta$=~\unit{75}{\nano\meter},~\unit{120}{\nano\meter} and,~\unit{200}{\nano\meter} is the Permalloy thickness. The Py thickness was calibrated by x-ray reflectivity. We estimate that the damping enhancement due to spin pumping into the Ta layer is two orders of magnitude smaller than the intrinsic damping of the Permalloy layer for Permalloy samples of these thicknesses. 
The various combinations of capping and seed layers are chosen to determine the sensitivity of our results on the spinwave boundary conditions and the resultant mode profiles. Prior to deposition, the substrates are cleaned by Ar plasma sputtering. The samples are coated with approximately \unit{150}{\nano\meter} of PMMA in order to avoid electrical shorting when samples are placed directly on the CPW. The CPW has a center conductor width of $\w = \unit{100}{\micro\meter}$. The SWR are characterized using field-swept VNA-FMR \cite{Ding2004,Kalarickal2006,Neudecker2006} in the out-of-plane geometry (see Fig.~\ref{fig:rad}) with an external static magnetic field $H_{0}$ applied perpendicular to the sample plane. The excitation microwave field $h^x(x,y)$ is applied over a frequency range of \unit{10}{\giga\hertz} to \unit{30}{\giga\hertz}. A VNA is used to measure the complex $S_{21}$ transmission parameter (ratio of voltage applied at one end of the CPW to voltage measured at the other end) for the waveguide/sample combination. The change in $S_{21}$ due to the FMR of the sample is then fitted with a linear superposition of complex susceptibility tensor components $\chi_{n}$,
\begin{align}
\begin{split}
\Sto(H_{0})&=\sum^{N}_{n = 0}A_{n} \chi_{n}(H_{0}) e^{i\phi_{n}} + \text{linear~background}\\
\end{split}
\end{align}
with the mode number $n$, phase $\phi_{n}$, and dimensionless mode amplitude $A_{n}$, as defined in the Appendix. A complex linear background and offset is included in the fit. The susceptibility components are derived from the Landau-Lifshitz equation for the perpendicular geometry; in the fixed-frequency, swept-field configuration, we obtain~\cite{Nembach2011}
\begin{align}
\begin{split} \chi(H_{0})_{n}&=\frac{\Ms(H_{0}-M^{\text{eff}}_{n}-\Hex)}{(H_{0}-M^{\text{eff}}_{n}-\Hex)^{2}-\left(H^{\text{eff}}\right)^{2}-i\Delta H_{n} (H_{0}-M^{\text{eff}}_{n}-\Hex)}
\end{split}
\end{align}
with $H^{\text{eff}}=\omega/(\gamma\mu_{0})$ and $M^{\text{eff}}_{n}=M_{\text{s}}-H_{k}$, where $H_{\text{k}}$ is the perpendicular anisotropy, \Hex is the exchange field (defined below), and $\Delta H_{n}$ is the linewidth. An example of the resulting fits for the complex $S_{21}$ data is shown in Fig.~\ref{spectrum}~(a)~and~(b).

\section{Experiment} 

We detect both even and odd PSSW modes. If we assume a uniform excitation field and Dirichlet boundary conditions (completely pinned), only odd modes would be detected. Alternatively if we assume Neumann boundary conditions (completely unpinned), only the fundamental mode would be detected. 

Two effects can contribute to our ability to detect all the PSSW modes. First, the excitation field profile might not be uniform due to eddy current shielding\cite{Kostylev2009,Bailleul2013}. Second, the interfacial boundary conditions might be asymmetrical, as alluded to above. According to the criterion in Ref.~[$_{\cite{Bailleul2013}}$], the threshold sheet resistance for the onset of eddy current shielding at \unit{20}{\giga\hertz} is $\unit{0.065}{\ohm\per \square}$. We estimate that the sheet resistance for our \unit{200}{\nano\meter} is in excess of $\unit{0.345}{\ohm\per \square}$, so we conclude that the eddy current shielding is relatively weak for our samples.


On the other hand, all modes are in principle detectable if we assume asymmetric interfacial anisotropy. For the sake of simplicity of the analysis, we will assume interfacial anisotropy for a single interface, then use an optimization approach to determine the wavenumber of the modes that is consistent with such a hypothesis. However we must emphasize that this approach does not provide a unique fit for the measured distribution of resonance fields for the PSSW spectrum, but simply allows us to accommodate for the wavenumber values required to be consistent with the measured spectrum. As such, the fitted value for $K_{\text{s}}$ is to be interpreted as no more than a self-consistent value associated with only one of many possible scenarios.


If we assume negligible magnetocrystalline perpendicular anisotropy $H_{\text{k}}$, $\Hres$ is related to the exchange field via 
\begin{align}
\begin{split} 
\Hres &= \Hex + \Ms, \\ \text{with } \Hex &= \frac{2A_{\text{ex}}}{\mu_{0}\Ms} k_{n}^{2}.
\label{hex}
\end{split} 
\end{align}
Here, $k_{n}$ is the spinwave wavevector, and $A_{\text{ex}}$ is the exchange energy that is related to the spinwave stiffness $D$ via $D=\frac{2A_{\text{ex}}g\mu_{\text{B}}}{\Ms}$. 
On the other hand, if we want to include interfacial anisotropy for a single interface in our analysis, we can numerically solve the transcendental equation~\cite{Soohoo1963}
\begin{equation}
\left(-\frac{1}{2}k_{n}a+\frac{K_{\text{s}}}{2A_{\text{ex}}k_{n}} +1\right) \text{tan}(k_{n}\delta)=\frac{K_{\text{s}}}{2A_{\text{ex}}k_{n}}\;,
\label{soohoo}
\end{equation}
where $K_{\text{s}}$ is the interfacial anisotropy, and $a=\unit{0.3547}{\nano\meter}$ is the lattice constant~\cite{Mathias2012}. 
We minimize the residue of the fit of Eq.~\eqref{hex} to $\Hres$ with the fitting parameters $\Ms$, $A_{\text{ex}}$, and $K_{\text{s}}$ from Eq.~\eqref{soohoo} by use of a Levenberg-Marquardt optimization algorithm. This yields the pairs ($k_{n}$,\Hex) shown in Fig.\ref{spectrum}~(c) for all modes.

From the fit, we obtain a saturation magnetization of $\mu_{0}M_{\text{s}}= \unit{1.02\pm0.01}{\tesla}$, in agreement with that determined by magnetometry. The exchange stiffness constant of $D = \unit{3.22\pm0.04}{\milli\electronvolt\nano\meter^{2}}$ is close to a value of $D \approx \unit{3.1}{\milli\electronvolt\nano\meter^{2}}$ reported by \textit{Maeda~et~al.}~\cite{Maeda1973}.


The exchange fit also yields a single surface anisotropy $K_{\text{s}}$ that depends on the cap and seed layer configurations. For the Ta(3)-Py($\delta$)-$\text{Si}_{3}\text{N}_{4}$(3) sample series, $K_{\text{s}}=\unit{(5.1\pm0.8)\times10^{-4}}{\joule\per\meter^{2}}$, while all the other samples have a higher $K_{\text{s}}$ of \unit{(7\pm1)\times10^{-4}}{\joule\per\meter^{2}}. All values for $K_{\text{s}}$ are in the range of other reported interface anisotropies for Permalloy layers of these thicknesses \cite{Bailey1973}.\\

\begin{figure}[!htp]
\begin{center}
		\includegraphics [width=0.8\textwidth] {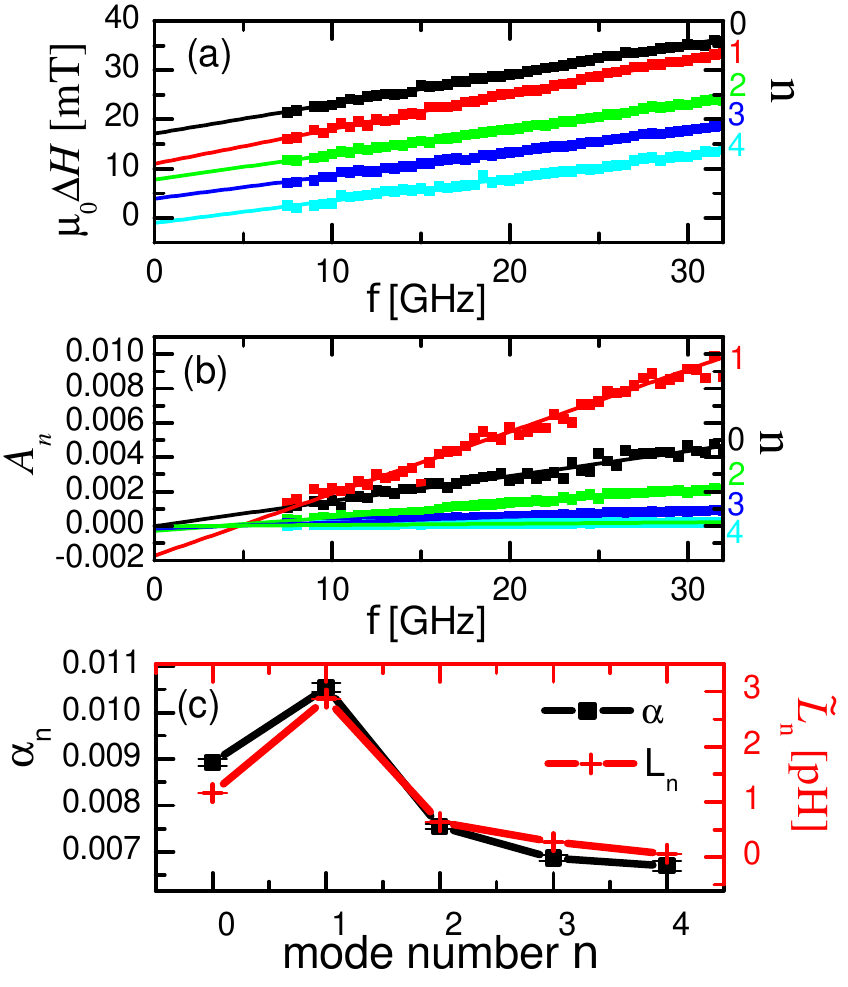}
		\caption{Parameter extraction for the first 5 PSSWs 
of the Ta(3)-Py(200)-$\text{Si}_{3}\text{N}_{4}$(3) sample. (a) Extraction of $\alpha$ from the linewidth $\mu_{0}\Delta H$ (data points) via linear fits (lines); staggered for display. (b) Extraction of the normalized mode inductance $\Ln$ (data points) from the resonance amplitude $\A(f)$ via linear fits (lines). (c) $\alpha_{n}$ (black squares) and $\Ln$ (red crosses) for each PSSW.}
		
		\label{fig:avsb}
\end{center}
\end{figure}

We now turn to the linewidth $\Delta H_{n}$ and the amplitude $\A$ for the individual modes. The Gilbert damping parameter $\alpha_{n}$ is extracted from the slope of the linewidth vs. frequency $f$ plot~\cite{Nembach2013} shown in Fig.~\ref{fig:avsb}(a) via 
\begin{equation}
\Delta H_{n}= \frac{4\pi\alpha_{n} f}{|\gamma|\mu_{0}}+\Delta H^{0}_{n},
\label{eq:deltaH}
\end{equation}
where $\Delta H^{0}_{n}$ is the inhomogeneous broadening that gives rise to a nonzero linewidth in the limit of zero frequency excitation.
The normalized inductance of the modes $\Ln$ is extracted in a similar fashion from the dependence of the mode amplitude $\A$ on the frequency $f$, see Fig.~\ref{fig:avsb}(b) and Eq.~\eqref{eq:an} in the Appendix
\begin{equation}
\A=2\pi f\frac{\Ln}{Z_{0}}+A_{n}^{0},
\label{Z}
\end{equation} 
where $A_{n}^{0}$ is an offset for each mode. $A_{n}^{0}$ is a phenomenological fitting parameter, which is not yet fully understood. 

We plot $\alpha_{n}$ and $\Ln$ as a function of mode number $n$ in Fig.~\ref{fig:avsb} (c). The damping and the normalized mode inductance are found to be proportional.
\begin{figure}[!htp]
\begin{center}
		\includegraphics [width=0.8\textwidth] {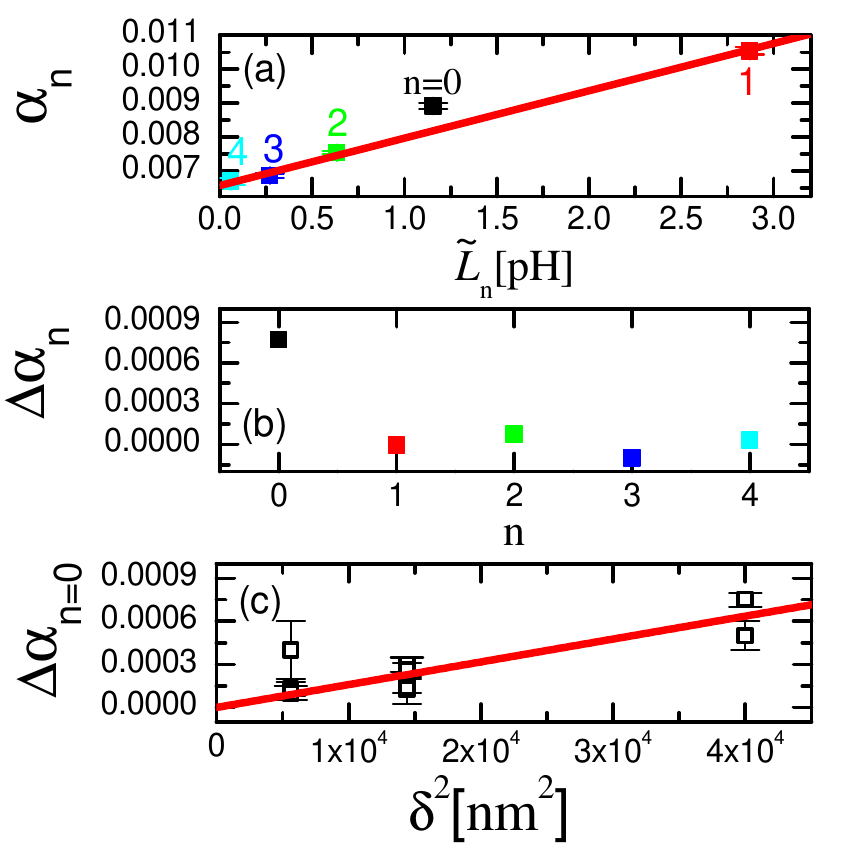}
		\caption{Damping $\alpha_{n}$ and inductance $\Ln$ for the Ta(3)-Py(200)-$\text{Si}_{3}\text{N}_{4}$(3) sample. (a)  Linear fit of $\alpha$ to Eq.~\eqref{rad} where the fit is constrained to the n=~1,~2,~3,~4 modes. (b) The residual of the linear fit, showing enhanced damping for the 0-th order mode (black). We attribute the enhanced damping to an eddy current contribution. (c) Enhanced 0-th order mode damping for all samples. The red line is a fit of the data points to the eddy current damping model from Eq.~\eqref{eddy}.}		\label{fig:da}
\end{center}
\end{figure}
In order to explore this correlation, we plot $\alpha_{n}$ vs. $\Ln$ in Fig.~\ref{fig:da}(a). Here, the data for $\alpha_{n}$ vs. $\Ln$ are linearly correlated for all modes except for $n=0$, as seen by the linear fit (line) to the data for $n\geq1$. This is as expected for the radiative damping model, as summarized in Eq.~\eqref{rad}. The additional damping of the fundamental mode is interpreted as the result of eddy current damping, as quantified in Eq.~\eqref{eddy}. In Fig.~\ref{fig:da}(b), we plot the residual $\Delta\alpha_{n}$ of the linear fit shown in Fig.~\ref{fig:da}(a) for all modes. $\Delta\alpha_{n}$ is negligible for all modes except for $n=0$. We extract $\Delta \alpha_{n=0}$ for all the samples and plot $\Delta \alpha_{n=0}$ vs. $\delta^{2}$ in Fig.~\ref{fig:da}(c). 

It appears that $\Delta \alpha_{n=0}$ for all the samples scales linearly with $\delta^{2}$, as expected from Eq.~\eqref{eddy} for eddy current damping. Simultaneous weighted fits of all the data to Eq.~\eqref{eddy} yields $C=0.4\pm0.1$. This value suggests a localization of eddy currents, since $C$ corrects for the eddy current distribution in the sample.


For the $n\geq1$ modes, it can be shown\cite{Pincus1960} that $\alpha^{\text{eddy}}_{n}\propto 1/k_{n}^{2}$. The calculated wavevectors from Eq.~\eqref{soohoo} for the $n=1$ mode of all the samples is at least a factor three larger than that of the $n=0$ mode and, therefore, the eddy current damping of the $n=1$ mode is predicted to be approximately one order of magnitude smaller than the eddy current damping of the $n=0$ mode. Thus, the eddy current damping of the $n\geq1$ modes is negligible to within the error bars, i.e., $\alphae_{n}\approx 0$ for $n\geq 1$. This supports the analysis of the data in Ref.~$[_{\cite{Li2014}}]$, which also neglects the eddy current damping in higher order modes.

\section{Extraction of the radiative contribution to the damping}

\begin{figure}[!htp]
\begin{center}
		\includegraphics [width=0.8\textwidth] {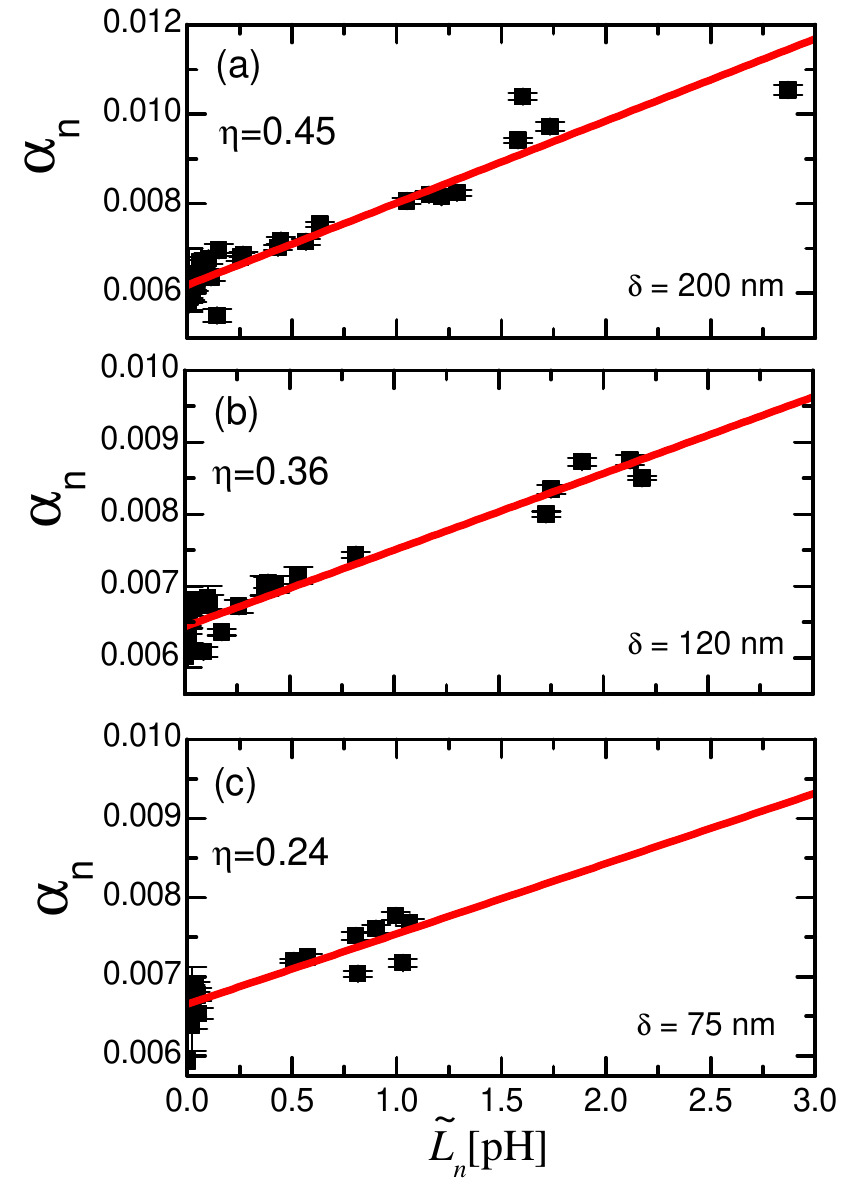}
		\caption{Dependence of damping $\alpha$ on the normalized mode inductance $\Ln$ after correction for the eddy current damping $\Delta\alpha_{0}$ of the fundamental mode, plotted for all sample configurations and different thicknesses: (a)~$\delta=\unit{200}{\nano\meter}$, (b)~$\delta=\unit{120}{\nano\meter}$ and (c) $\delta=\unit{75}{\nano\meter}$. The red lines are weighted linear fits to the data by use of Eq.~\eqref{rad}, that describes the radiative component of the damping.}				\label{fig:raddamp}
\end{center}
\end{figure}

By use of Eq.~\eqref{eddy} and our fitted value of $C=0.4$, we subtract the eddy current contribution to the damping of all the $n=0$ modes to obtain a corrected damping value $\alpha_{n=0}'$, where $\alpha_{n=0}'=\alpha_{n=0}-\alphae(C=0.4)$. The corrected data for all the modes are plotted in Fig.~\ref{fig:raddamp}.

Figures~\ref{fig:raddamp}~(a)~to~(c) group all data obtained for a set of samples with identical Py thickness $\delta$. The lines are linear fits to Eq.~\eqref{rad}. For each thickness $\delta$, we observe a significant correlation of $\alpha_{n}$ and $\Ln$ for all seed and cap layer configurations, as expected for a radiative damping mechanism.

Furthermore, by use of Eq.~\eqref{simplealpha} for the $n=0$ mode of the \unit{75}{\nano\meter} thick sample, using a value of $\eta\approx0.46$ as determined in the Appendix, we estimate $\alpha^{\text{rad}}_{0}\approx 0.00023$

The experimentally determined value is $\alpha^{\text{rad}}_{0}\approx 0.00035\pm 0.0001$. The deviance from the calculated value is possibly due to non-uniformities of both the excitation field and magnetization profile in Eq.~\eqref{simplealpha}, that requires the solution of the integral in Eq.~\eqref{eq:Kgen}. 
Nevertheless the estimated value for $\alpha_{0}^{\text{rad}}$ is of the correct order of magnitude.\\

\begin{figure}[!htp]
\begin{center}
		\includegraphics [width=0.8\textwidth] {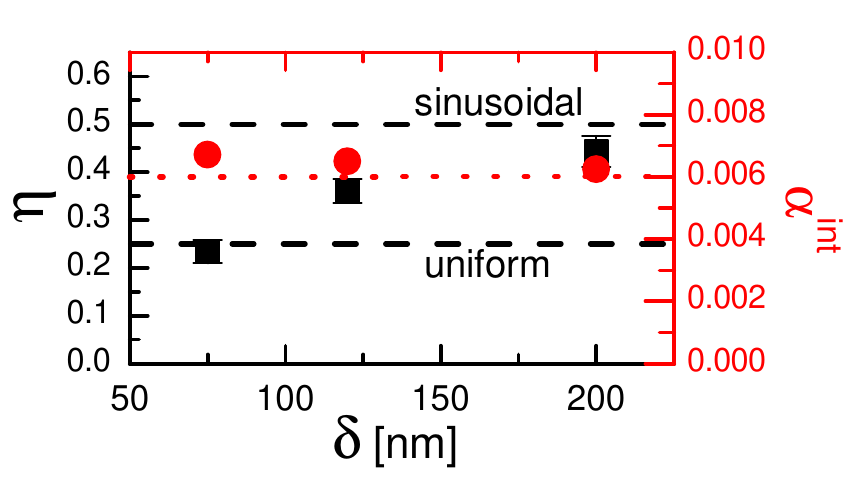}
		\caption{Mode profile parameter $\eta$ (black squares, left axis) and intrinsic damping $\alpha_{\text{int}}$ (red circles, right axis) as a function of Py thickness $\delta$. The mode profile parameter $\eta$ lies between the value of $1/2$ for sinusoidal PSSWs with anti-symmetric boundary conditions (dashed black line) and the value of $1/4$  for the uniform mode (both dashed black lines). 
The intrinsic damping is close to $\alpha_{\text{int}}=0.006$ (dotted red line).}		
		\label{fig:summary}
\end{center}
\end{figure}

We determine the intrinsic damping $\alpha^{\text{int}}$ from the $\Ln=0$ intercept of the linear fits in Fig.~\ref{fig:raddamp}. We plot $\alpha^{\text{int}}$ for the three values of $\delta$ in Fig.~\ref{fig:summary}~(right scale). We find that $\alpha^{\text{int}}$ is approximately constant to within~$\pm 5\%$ for all samples. In addition the average value over all the film thicknesses is in reasonable agreement to the previously reported value of $\alpha^{\text{int}}$= 0.006 (dotted red line)\cite{Shaw2009}. 

The other fitting parameter $\eta$, extracted from the slope of $\alpha_{n}$ vs. $\Ln$, is also plotted in Fig.~\ref{fig:summary}~(left scale). For anti-symmetric boundary conditions, $\eta=1/2$ is expected, whereas for the uniform mode, $\eta=1/4$.\\
 We see that the fitted values lie exclusively within these extremes, within error bars.\\
%

\begin{figure}[!htp]
\begin{center}
		\includegraphics [width=0.8\textwidth] {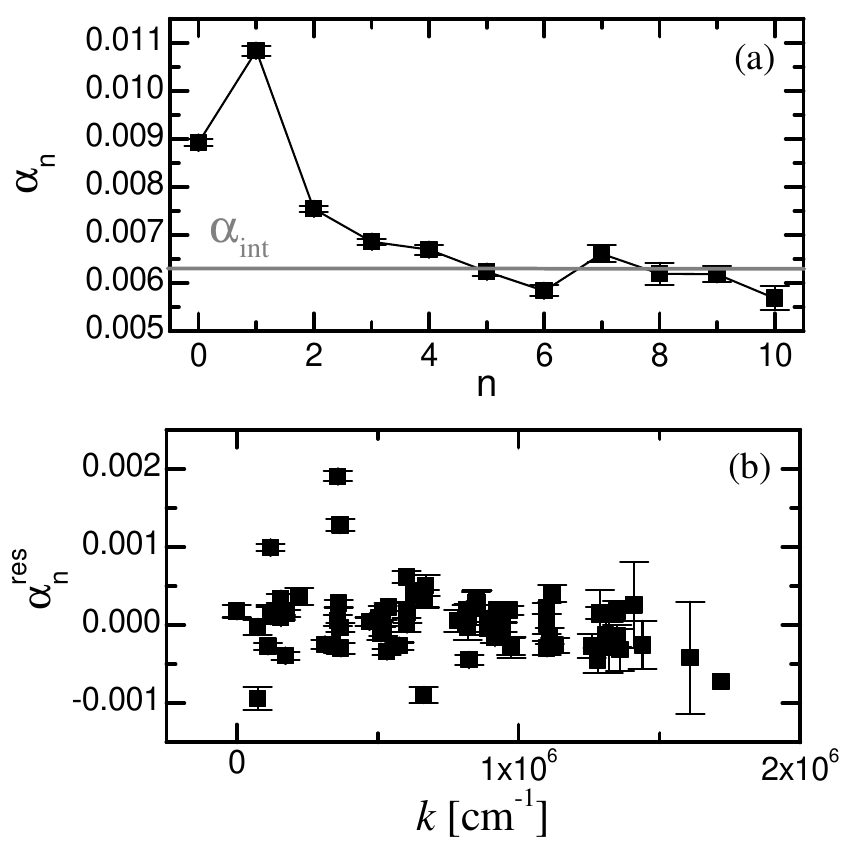}
		\caption{(a) The measured damping for the first 11 PSSW modes of the  Ta(3)-Py(200)-$\text{Si}_{3}\text{N}_{4}$(3) sample. The enhanced damping due to inductive coupling to the waveguide and eddy currents in the sample only affects the first five modes at wavevectors $\leq \unit{7\times10^{5}}{\centi\meter^{-1}}$. (b) The residual damping for all detected modes for all samples is plotted against their respective wave vector $k$. Within the scatter, no dependence of the residual damping on $k$ is observed.}
		\label{resalpha}
\end{center}
\end{figure}

There have been recent reports of a non-zero, wavenumber-dependent component for damping for both localized eigenmodes in magnetic nanostructures\cite{Nembach2013} and PSSWs in thick Permalloy films\cite{Li2014}. Such exchange-mediated damping of the form $\alpha^{\text{ex}}:=A_{\text{ex}}k^{2}$ was originally predicted by Bar$'$yakhtar based on symmetry alone\cite{V.Baryakhtar1986}. Nembach, et al.\cite{Nembach2013}, obtained a value of $A_{\text{ex}}=\unit{1.4}{\nano\meter^{-2}}$, whereas Li, et al.\cite{V.Baryakhtar1986}, found a much smaller value of \unit{0.09}{\nano\meter^{-2}}. To determine whether wavenumber-dependent damping is apparent in our data, we examined the residual damping after subtraction of both the intrinsic damping \alphaint and the radiative damping \alphar from all the modes, as well as subtraction of the eddy current damping from the $n=0$ mode. The residual damping $\alpha^{\text{res}}$ is plotted in Fig.~\ref{resalpha}~(b). Within the scatter of $\cong \pm0.001$, $\alpha^{\text{res}}$ does not have any clear dependence on $k$. Thus, we obtain an upper bound of $A_{\text{ex}}\leq\unit{0.045}{\nano\meter^{-2}} $ for this particular system, given the sensitivity of our measurements. For comparison and to ensure that the subtraction of \alphaint, \alphar, and \alphae did not hide a potential $k^{2}$ contribution the measured damping of the Ta(3)-Py(200)-$\text{Si}_{3}\text{N}_{4}$(3) sample up to the $n=10$ mode is shown in Fig.~\ref{resalpha}~(a). For $n\geq5$ the measured damping scatters around the for the \unit{200}{\nano\meter} samples determined intrinsic damping \alphaint and no trend for higher mode numbers (larger $k$ values) is discernible.

Tserkovnyak, et al., calculated the damping coefficient $A_{\text{ex}}$ in terms of a microscopic model for the diffusive transport of dissipative transverse spin current within a ferromagnetic metal\cite{Tserkovnyak2002}. The theory in Ref.~[$_{\cite{Tserkovnyak2002}}$] framed the exchange-mediated damping in terms of a so-called transverse spin conductivity $\sigma_{\bot}$,
\begin{equation}
A_{\text{ex}}=\left(\frac{\gamma}{\Ms} \right)\left(\frac{\hbar}{2e} \right)^{2} \sigma_{\bot},
\end{equation}
where
\begin{equation}
\sigma_{\bot}:=\left(\frac{\sigma}{\tau} \right)\left(\frac{\tau_{\bot}}{\left(1+\left(\omega_{\text{ex}}\tau_{\bot} \right)^{2} \right)} \right),
\end{equation}

with the exchange splitting $\hbar \omega_{\text{ex}}$, the conductivity $\sigma$, the spin scattering time $\tau$, and transverse spin scattering time $\tau_{\bot}$. Given that $\hbar \omega_{\text{ex}}\approx \unit{1}{e\volt} $ for Permalloy, the maximum value for $A_{\text{ex}}$ predicted by the transverse spin current theory is \unit{0.001}{\nano\meter^{-2}}. Insofar as we are not able to observe any such wavenumber-dependent damping down to the level of \unit{0.045}{\nano\meter^{-2}}, our results are consistent with the predictions of the microscopic theory.

While the theory in Ref.~[$_{\cite{Li2014}}$] is specific to the microscopic mechanism of transverse spin accumulation in a metallic ferromagnet, the phenomenology of exchange mediated damping, as described in Ref.~[$_{\cite{V.Baryakhtar1986}}$], is not limited to such a microscopic mechanism. As such, it remains plausible that extrinsic material-specific parameters that have not yet been identified could be responsible for the previously reported values for $k^2$ damping.
For example the presence of anti-symmetric exchange at interfaces, i.e. the Dzyaloshinskii-Moriya interaction (DMI) could enhance the coupling between magnons and Stoner-excitations insofar as the DMI gives rise to exotic spin textures\cite{Romming2013} with nanometer length scales, that are comparable to the wavelength of low energy Stoner-excitations\cite{Ibach2006}. Thus, the results of Ref.~[$_{\cite{Nembach2013}}$] could be a manifestation of interfacial enhancement for $\A^{\text{ex}}$, insofar as the magnetic films used in Ref.~[$_{\cite{Nembach2013}}$] are only \unit{10}{\nano\meter} thick.\\

\begin{figure}[!htp]
\begin{center}
		\includegraphics [width=0.7\textwidth] {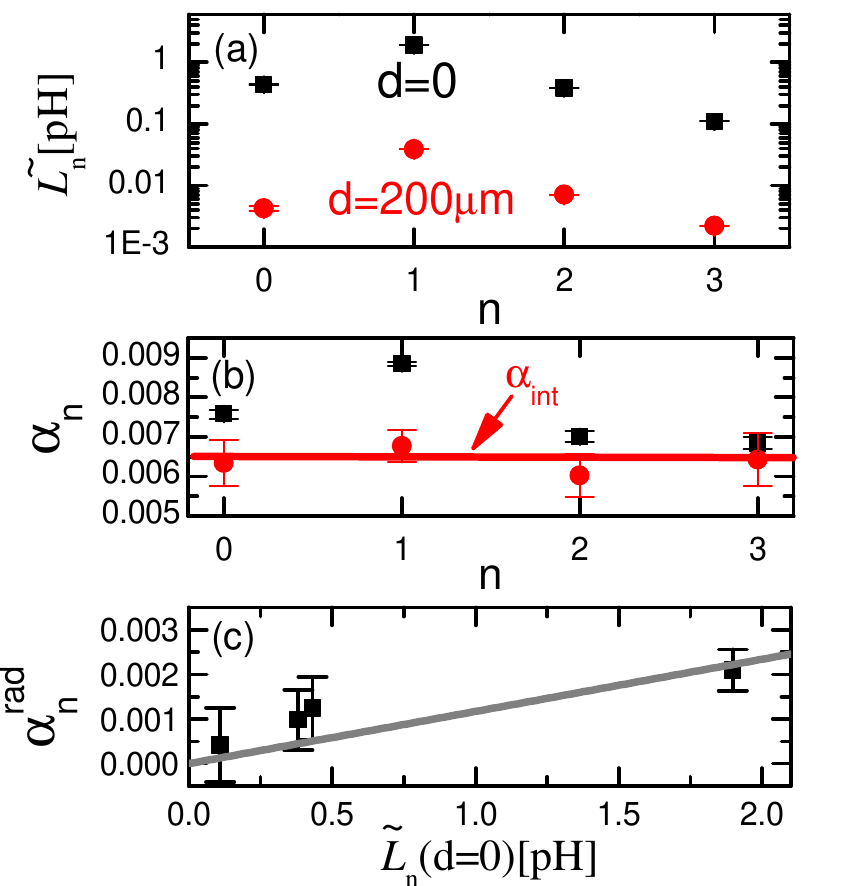}
		\caption{Measurement of the first four PSSWs of the Ta(3)-Py(120) sample with and without a spacer inserted between sample and CPW. (a) Inductance $\Ln$ determined for the sample directly on the CPW (black squares) and for a \unit{200}{\micro\meter} spacer between sample and CPW (red circles). (b) The resulting damping constants for both measurements (same symbols and colors). The red line is the previously extracted intrinsic damping $\alpha^{\text{int}}$. (c) The difference between the damping with and without the spacer (black squares) is in good agreement with the radiative damping from Fig.~\ref{fig:raddamp}(b) (gray line).}	
		\label{fig:test}
\end{center}
\end{figure}

In another experiment, we further validate the presence of radiative damping and demonstrate an alternative method to determine  $\alphar$ by varying the distance $d$ in Eq.~\eqref{kquist} between the sample and waveguide. To this end, we insert a $d = \unit{200}{\micro\meter}$ glass spacer between the sample and waveguide. By comparing $h(0,0)$ to $h(0,\unit{200}{\micro\meter})$ via Eq.~\eqref{kquist}, we estimate that the insertion of the spacer decreases the microwave magnetic field by about a factor of 6.25. Referring to Eq.~\eqref{z}, the normalized mode inductance $\Ln$ decreases by a factor of $\cong40$. To determine the effect of the reduced inductive coupling on the radiative damping, we used VNA-FMR to measure the first 4 modes for the Ta(3)-Py(120) sample with and without the spacer. The effect of the spacer can be seen in the raw data, reducing the linewidth of the first two modes measured at \unit{10}{\giga\hertz} in the \unit{120}{\nano\meter} samples by approximately 6~Oe, well outside error bars. The fitted values of $\Ln$ are shown in Fig.~\ref{fig:test}~(a). Indeed, $\Ln$ decreases on average for all modes by a factor of $\cong50$ after inserting the spacer, in good agreement with the predictions of Eq.~\eqref{kquist} and~\eqref{z}. Thus, we will assume that $\alphar$ is negligible when the spacer is used. The data for the damping $\alpha_{n}$ of the first four modes, both with and without the spacer, are plotted in Fig.~\ref{fig:test}~(b). Indeed, the damping determined from the measurement with the spacer layer (circles) is consistently lower than that found without the spacer layer (squares). The line in Fig.~\ref{fig:test}~(b) is the previously determined intrinsic damping. Under the assumption that the radiative damping contribution is given by $\alphar=\alpha_{n}(d=0)-\alpha_{n}(d=\unit{200}{\micro\meter})$, we plot $\alphar$ vs $\Ln(d=0)$ in Fig.~\ref{fig:test}~(c). The line is the calculated $\alphar$, where we used Eq.~\eqref{rad} with $\eta=0.35$ and $\delta=\unit{120}{\nano\meter}$, as determined from the fits in Fig.~\ref{fig:raddamp}. Good agreement between the calculated and measured values for \alphar are obtained, which demonstrates the self-consistency of our analysis. Of great importance is that the spacer-layer approach can also be used to determine the radiative contribution to the damping in the absence of PSSWs (single resonance). By measuring $\alpha$ for varying distance $d$ between sample and waveguide and extrapolating $\alpha$ to $d \rightarrow \infty$, both the intrinsic value for the damping and the radiative contribution can be determined, under conditions where eddy current damping is negligible.

\section{Summary}
In summary, we identified three contributions to the damping in PSSWs: Intrinsic damping \alphaint, eddy current damping \alphae, and radiative damping $\alphar$. The latter exhibits a linear dependence on the normalized sample inductance $\Ln$ in a waveguide based FMR measurement. We attribute this linear dependence to radiative losses that stem from the inductive coupling between the sample and the waveguide. The radiative damping term is inherent to the measurement process and is thus present in all FMR measurements. The radiative damping constitutes up to 40~\% of the total damping of the spin wave modes in our \unit{200}{\nano\meter} thick Permalloy films. Furthermore, the radiative damping can be already important for much lower film thicknesses, in materials with small intrinsic damping.

As an example, the radiative damping calculated from Eq.~\eqref{simplealpha} for a \unit{20}{\nano\meter} thick and \unit{1}{\centi\meter} long sample of Yttrium-Iron-Garnet (YIG), measured on a \unit{100}{\micro\meter} wide wave guide is $\alpha^{\text{rad}}_{0}\approx 1.26\cdot10^{-4}$ . When compared to the reported value for the damping of $\alpha=2.3\cdot10^{-4}$,\cite{AllivyKelly2013} we see, that the radiative part of the damping, among others\cite{Schneider2005}, can substantially influence the determination of $\alpha_{\text{int}}$. As such, careful analysis of $\alpha$ vs. inductance is required to isolate the radiative damping contribution.

\section{Aknowledgement}
The authors are grateful for the assistance of Mikhail Kostylev in the development of our theoretical analysis.

\section{appendix}
\subsection{Derivation of \alphar}\label{deriverad}
In this section, we derive model equations for the normalized mode inductance $\Ln$, mode amplitude $\A$, and the radiative damping $\alphar$. We are restricting our analysis to the case of ideal perpendicular standing spin wave modes that only vary through the film thickness without any lateral variation. It is assumed that the excitations are in the perpendicular geometry with the magnetization saturated out of the film plane. As such, the response to the z-coordinate component of the microwave excitation field above the waveguide can be neglected. In addition, the magnetization precession is always circular. As such, the magnetization dynamics in the x- and y-coordinates in response to the microwave field generated by the waveguide are degenerate, outside of a phase factor of $\pi/2$. This eliminates the need to explicitly consider the full Polder susceptibility tensor in the calculation of the sample response to the excitation field. The sample dimensions are $l$ along the waveguide direction, $\delta$ in thickness, but infinite in the lateral direction.\\
We begin by introducing the concepts of a spin wave mode susceptibility $\chi_{n}$, and the dimensionless, normalized spin wave amplitude $q_{n}(z)$ for the nth spin wave mode, such that the magnetic excitation of amplitude in $x$-direction $\tilde{m}_{n}^{x}\left( {{H}_{0}},I;z \right)$ that results from the application of a microwave magnetic field of amplitude $\hx$, given in Eq.~\eqref{kquist}, driven by an ac current $I={{{V}_{\text{in}}}}/{{{Z}_{0}}}\;$ in an applied field ${{H}_{0}}$ is given by
\begin{equation}
\tilde{m}_{n}^{x}\left( {{H}_{0}},I;z \right):=\tilde{m}_{n}^{x}\left(z\right)={{q}_{n}}\left( z \right){{\chi }_{n}}\left( {{H}_{0}} \right)\left\langle {{q}_{n}}\left( z \right){{h}^{x}}\left( I;x,z \right) \right\rangle 
\label{eq:mtilde}
\end{equation} 

where the quantity in brackets is simply the overlap integral of the excitation field and the spatial profile of the nth spin wave mode. The magnetic excitation of amplitude in $y$-direction $\tilde{m}_{n}^{y}\left(z\right)$ can be written in a similar way. In the trivial case of a uniform excitation field and uniform spin wave mode, we recover the usual relation between the excitation field and the magnetization dynamics via the Polder susceptibility tensor component, $\chi^{xx}$. However, if the product of the mode profile and excitation field has odd spatial symmetry, dynamics are not excited, as we expect. The overlap integral is nothing more than the spatial average of the mode/excitation product:
\begin{equation}
\left\langle q_n(z) \hx \right\rangle=\frac{1}{\w\delta}\intx\intz q_{n}(z)\hx.
\label{2}
\end{equation}

First, the power transferred to the waveguide via inductive coupling with the spin wave dynamics is given by

\begin{equation}
{{P}_{n}}=\frac{{{\left| {{\partial }_{t}}{{\Phi }_{n}}\left( {{H}_{0}},I \right) \right|}^{2}}}{2{{Z}_{0}}},
\end{equation}

where

\begin{equation}
{{\partial }_{t}}{{\Phi }_{n}}\left( {{H}_{0}},I \right)={{\mu }_{0}}\ell \int\limits_{-\infty }^{\infty }{dx}\int\limits_{d}^{\delta +d}{dz}\left( {{\partial }_{t}}m_{n}^{x}\left(z\right)\right){{\tilde{h}}^{x}}\left( x,z \right),
\label{flux}
\end{equation}

with $\tilde{h}^{x}\left( x,z \right)=\hx/I$.\\
It is important to recognize at this point that the power dissipation is not constant with time, given that ${{P}_{n}}$ is proportional only to ${{\partial}_{t}}m_{n}^{x}$. As such, the damping associated with the re-radiation of the microwave energy back into the waveguide is best characterized with an anisotropic damping tensor, to be elaborated upon more fully later in this Appendix.
To calculate the energy of the spin wave mode, we start by defining a spatially averaged spin wave excitation density\cite{L'vov1994}, 

\begin{equation}
\left\langle \mu _{n}^{2}\left( {{H}_{0}},I \right) \right\rangle =\frac{\intx \intz \left[\left(\partial_{t}m_{n}^{x}\left(z\right)\right)\left(m_{n}^{y}\left(z\right)\right)^{*}-\left(\partial_{t}m_{n}^{y}\left(z\right)\right)\left(m_{n}^{x}\left(z\right)\right)^{*}\right]}{4\omega\delta \w}.
\end{equation}

We can then calculate the magnon density ${{N}_{n}}$ associated with the nth spin wave excitation as

\begin{equation}
{{N}_{n}}=\frac{\left\langle \mu _{n}^{2}\left( {{H}_{0}},I \right) \right\rangle }{2g{{\mu }_{B}}{{M}_{s}}}
\end{equation}

The total energy associated with the spin wave mode is given by

\begin{equation}
{{E}_{n}}=\frac{\omega \left\langle \mu _{n}^{2}\left( {{H}_{0}},I \right) \right\rangle }{\gamma {{M}_{s}}}\delta \ell \w
\end{equation}


The energy dissipation rate ${{\left( {1}/{{{T}_{1}}}\right)}_{n}}$ for the nth mode is therefore

\begin{equation}
{{\left( \frac{1}{{{T}_{1}}} \right)}_{n}}=\frac{{{P}_{n}}}{{{E}_{n}}}=\frac{2\mu _{0}^{{}}\ell {{\omega }_{M}}}{{{Z}_{0}}}\frac{{{\left| \int\limits_{-\infty }^{\infty }{dx}\int\limits_{d}^{\delta +d}{dz}\left( {{\partial }_{t}}m_{n}^{x}\left(z\right) \right){{{\tilde{h}}}^{x}}\left( x,z \right) \right|}^{2}}}{\int\limits_{-\infty }^{\infty }{dx\int\limits_{d}^{\delta +d}{dz\left[{\left( {{\partial }_{t}}m_{n}^{x}\left(z\right)\right) \left( m_{n}^{y}\left(z\right)\right)^{*}}-\left( {{\partial }_{t}}m_{n}^{y}\left(z\right)\right) \left( m_{n}^{x}\left(z\right)\right)^{*}\right]}}},
\end{equation}

where ${{\omega }_{M}}=\gamma {{\mu }_{0}}{{M}_{s}}$. We then apply the Fourier transform to move into the frequency domain, where ${{\partial }_{t}}m_{n}^{x}\left( {{H}_{0}},I;z \right)\leftrightarrow i\omega \tilde{m}_{n}^{x}\left( {{H}_{0}},I;z \right)$, such that the energy relaxation rate $\left({1}/{{{T}_{1}}}\right)_{n}^{x}$ for magnetization oscillations along the x-axis is

\begin{equation}
\left(\frac{1}{{{T}_{1}}} \right)_{n}^{x}=\frac{\omega \mu _{0}^{{}}\ell {{\omega }_{M}}}{{{Z}_{0}}}{{K}_{n}},
\end{equation}

where

\begin{equation}
{{K}_{n}}:= \frac{{{\left| \int\limits_{-\infty }^{\infty }{dx}\int\limits_{d}^{\delta +d}{dz}\left( \tilde{m}_{n}^{x}\left(z\right) \right){{{\tilde{h}}}^{x}}\left( x,z \right) \right|}^{2}}}{\int\limits_{-\infty }^{\infty }{dx\int\limits_{d}^{\delta +d}{dz{{ \text{Im}\left[\tilde{m}_{n}^{x}\left(z\right) \left(\tilde{m}_{n}^{x}\left(z\right) \right)^{*} \right] }}}}}
\label{eq:Kn}
\end{equation}

is a dimensionless inductive coupling parameter. In the limiting case of the $n=0$ (i.e., uniform) mode with a uniform excitation field due to current flowing only through the waveguide center conductor, and an infinitesimal spacing between the waveguide and the sample, we have ${{K}_{0}}={\delta}/{4w}$. Substituting Eq.~\eqref{eq:mtilde} into Eq.~\eqref{eq:Kn}, we obtain the general result

\begin{equation}
{{K}_{n}}=\frac{{{\left| \int\limits_{-\infty }^{\infty }{dx}\int\limits_{d}^{\delta +d}{dz}{{q}_{n}}\left( z \right){{{\tilde{h}}}^{x}}\left( x,z \right) \right|}^{2}}}{\epsilon\int\limits_{d}^{\delta +d}{dz{{\left| {{q}_{n}}\left( z \right) \right|}^{2}}}},
\label{eq:Kgen}
\end{equation}

with $\epsilon = \left|\tilde{m}_{n}^{z} \right|/\left|\tilde{m}_{n}^{x} \right|$.\\
Since the energy dissipation rate for the case of radiative damping is anisotropic, it must be generally treated in the damping tensor formalism, where the Gilbert damping torque $\vec{T}$ is given by

\begin{equation}
{{T}_{k}}={{\varepsilon }_{ijk}}{{\alpha }_{ij}}{{\hat{m}}_{i}}{{\left( {{\partial }_{t}}\hat{m} \right)}_{j}}.
\end{equation}

The equation of motion is

\begin{equation}
{{\partial }_{t}}\hat{m}=-\gamma {{\mu }_{0}}\hat{m}\times \vec{H}+\vec{T}
\end{equation}

and $\hat{m}={{\vec{M}}}/{{{M}_{s}}}$ is the normalized magnetization. For the coordinates in Fig~\ref{fig:rad}, the only nonzero radiative damping tensor components are ${{\alpha}_{zx}}$ and ${{\alpha}_{yx}}$.
For the perpendicular FMR geometry, the relationship between the energy relaxation rate and the Gilbert damping components is

\begin{equation}
{{\left( \frac{1}{{{T}_{1}}} \right)}^{x}}={{\alpha }_{zx}}{{\omega }_{x}},
\end{equation}
and

\begin{equation}
{{\left( \frac{1}{{{T}_{1}}} \right)}^{y}}={{\alpha }_{zy}}{{\omega }_{y}},
\end{equation}

where $\omega_{x}$ and $\omega_{y}$ are the respective stiffness frequencies, defined as

\begin{equation}
{{\omega}_{i}}:= \frac{\gamma}{{{M}_{s}}}\frac{{{\partial }^{2}}{{U}_{m}}}{\partial {{{\hat{m}}}_{i}}}
\label{deltaomega}
\end{equation}

and ${{U}_{m}}$ is the magnetic free energy function. The frequency-swept linewidth $\Delta\omega=\gamma{{\mu }_{0}}\Delta H$, where $\Delta H$ is the field-swept linewidth in Eq.~\eqref{eq:deltaH}, is given by

\begin{align}
  & \Delta \omega =\frac{{{\left( \frac{1}{{{T}_{1}}} \right)}^{x}}+{{\left( \frac{1}{{{T}_{1}}} \right)}^{y}}}{2} \\ 
 & ={{\alpha }_{zx}}{{\omega }_{x}}+{{\alpha }_{zy}}{{\omega }_{y}}  
\end{align}

For perpendicular FMR, ${{\omega }_{x}}={{\omega }_{y}}=\omega $, and the specific case of anisotropic radiative damping, ${{\alpha }_{zx}}=\alpha _{n}^{\text{rad}}$, ${{\alpha }_{zy}}=0$, and we obtain

\begin{equation}
\alphar=\frac{1}{2\omega}\left( \frac{1}{T_{1}}\right)_{n}^{x}= \frac{\mu_{0}l\omega_{M}}{2Z_{0}}K_{n} 
\label{eq:arad0}
\end{equation}

and $\Delta \omega _{n}^{\text{rad}}=\alpha_{n}^{\text{rad}}\omega$. This is in contrast to the case of isotropic damping processes, such as eddy currents and intrinsic damping, where we obtain $\Delta \omega _{n}^{\text{iso}}=2\alpha _{n}^{\text{iso}}\omega$ instead. Thus, the net damping due to the sum of anisotropic radiative damping, and any other isotropic processes, is given by
\begin{equation}
{{\alpha }_{n}}={{\alpha }^{\text{int}}}+\alpha _{n}^{\text{eddy}}+\frac{\alpha _{n}^{\text{rad}}}{2}
\label{allalpha1}
\end{equation}
where ${{\alpha }_{n}}$ is the damping parameter in Eq.~\eqref{eq:deltaH} for the field-swept linewidth. 

We use a vector network analyzer (VNA) to measure the two-port $S$-parameter matrix element for the nth spin wave mode, $\Sto$. The matrix element is defined as the ratio of the voltage induced in the waveguide by the nth spin wave mode $V_{n}(H_{0})$ in an applied magnetic field $H_{0}$, and the excitation voltage \Vin,

\begin{equation}
\Sto := \frac{V_{n}(H_{0})}{V_{in}}.
\label{eq:S1}
\end{equation}

If we model the reactance of the nth spin wave mode as nothing more than a purely inductive element of inductance $L_{n}$ in series with an impedance matched transmission line, and if we assume the sample inductance is much smaller than the transmission line impedance, we can approximate $\Sto$ as

\begin{equation}
\Sto (H_{0}) \cong -\frac{i\omega L_{n}(H_{0})}{Z_{0}}, 
\label{eq:S2}
\end{equation}

where $L_{n}(H_{0})=\Phin/I$.\\

We define a normalized, field-independent mode-inductance \Ln as

\begin{equation}
\Ln := \frac{L_{n}(H_{0})}{\chi_{n}(H_{0})}, 
\label{eq:ln}
\end{equation}

and a dimensionless, field-independent mode-amplitude $\A$,
\begin{equation}
\A := \frac{i\omega \Ln}{Z_{0}}. 
\label{eq:an}
\end{equation}

such that 
\begin{equation}
\Sto (H_{0})=-\A\chi_{n}(H_{0}).
\end{equation}
Thus, \A is the dimensionless amplitude parameter that we obtain when fitting data for $\Sto (H_{0})$.
By use of Eqs.~\ref{flux},~\ref{eq:an},~and~\ref{eq:ln}, we can rewrite the mode-amplitude as 

\begin{equation}
\A := i\frac{\omega \mu_{0} l}{\w \delta Z_{0}}\left(\intx\intz q_{n}(z)\tilde{h^{x}}(x,z) \right)^{2}.
\label{eq:AAn}
\end{equation}

Remembering that the normalized mode inductance has a factor identical to the numerator of Eq.~\eqref{eq:Kgen}, we can rewrite the radiative damping in terms of the normalized mode inductance,

\begin{equation}
\frac{\alphar}{\Ln}=\frac{\omega_{M}\eta_{n}}{Z_{0}},
\label{eq:arad1}
\end{equation}

where

\begin{equation}
\eta_{n}:=\frac{\delta}{4\intz \left| q_{n}(z) \right|^{2}}
\label{eq:eta}
\end{equation}

We emphasize that Eq.~\eqref{eq:arad1} is a very general result, regardless of the details of the excitation field profile . Thus, even if the field profile is highly non-uniform due to the combination of eddy current and capacitive coupling effects\cite{Maksymov2013,Bailleul2013}, there should still be a fixed scaling between the radiative damping and the normalized inductance.\\
In the case of the uniform mode, $\eta=1/4$ and Eq.~\eqref{eq:arad1} reduces to

\begin{equation}
\frac{\alphar}{\Ln}=\frac{\omega_{M}}{4Z_{0}}.
\label{eq:arad2}
\end{equation}

However, for a sinusoidal mode of the form

\begin{equation}
q_{n}(z)=\text{cos}\left( \frac{\left(2n+1 \right)\pi z}{2\delta}\right)
\label{eq:cos}
\end{equation}

that is expected in the case of a pinned boundary condition at one interface and an open boundary condition at the other interface, as shown in Fig.~\ref{fig:rad}~(b), we obtain $\eta=1/2$ and

\begin{equation}
\frac{\alphar}{\Ln}=\frac{\omega_{M}}{2Z_{0}},
\label{eq:arad3}
\end{equation}

\begin{figure}[!htp]
\begin{center}
		\includegraphics[width=0.8\textwidth]{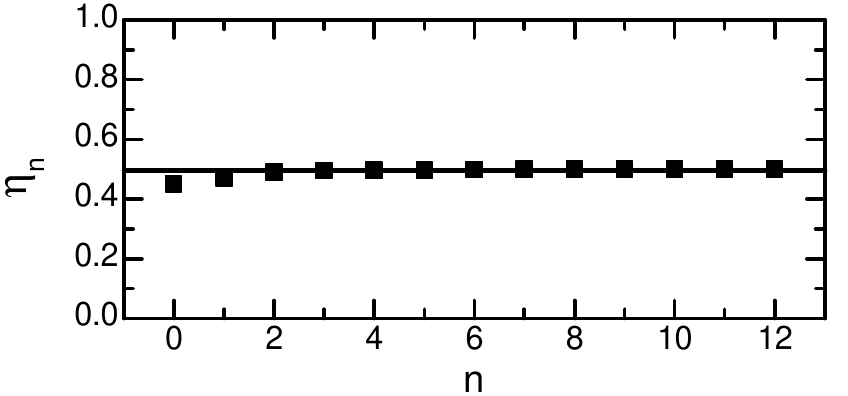}
		\caption{Dependence of calculated values for $\eta_{n}$ on mode number for the case of the spectral data presented in Fig.~\ref{spectrum}~(c)~and~\ref{fig:raddamp}~(a). Wavenumbers are extracted from those data via the procedure outlined in the main text of the paper, based upon a model with a single surface with interfacial anisotropy, and an unpinned boundary condition at the other interface. Within the context of that particular model, and the expected quadratic dependence of spin wave resonance frequency with wavenumber that it produces, we see that $\eta_{n}$ has a weak dependence on mode number, justifying our presumption that $\eta_{n}$ can be treated as a constant for the purposes of fitting the damping data.}	
		\label{fig:eta}
\end{center}
\end{figure}

For the case of the wavenumber values extracted from the data shown in Fig.~\ref{spectrum}~(c) for a \unit{200}{\nano\meter} Py film, we can determine the value for $\eta_{n}$ and the degree to which it can vary with mode number. We use the following form for the spin wave profile:

\begin{equation}
q_{n}(z)=\text{cos}\left( k_{n}z\right)
\label{eq:qn3}
\end{equation}

consistent with our assumption, when extracting $k_{n}$ from our PSSW data, that an unpinned boundary condition applies to only one of the interfaces, i.e. at $z=0$. Using these extracted values for the wavenumber, we obtain values for $\eta_{n}$ shown in Fig.~\ref{fig:eta}.

We see in Fig.~\ref{fig:eta} that the variation in $\eta_{n}$ with varying mode number is less than 10\%. Thus, to within first order, we can treat $\eta_{n}$ as a constant for the purposes of fitting our data, i.e. $\eta_{n}\cong\eta$.

\subsection{Derivation of \alphae} \label{eddyderivation}
For the derivation of the eddy current damping $\alpha_{\text{eddy}}$ uniform magnetization dynamics are assumed. The notation stays the same as for the radiative damping. 

Then the total flux passing through the magnetic film is 

\begin{equation}
{{\partial }_{t}}\Phi ={{\mu }_{0}}\ell {{\delta }}{{\left( {{\partial }_{t}}\vec{m} \right)}_{x}},
\end{equation}
where ${{\left( {{\partial }_{t}}\vec{m} \right)}_{x}}=\hat{x}\cdot {{\partial }_{t}}\vec{m}$.
The electrical power dissipated by the eddy-currents is

\begin{align}
  & {{P}_{\text{ind}}}=\frac{1}{2}\frac{{{\left| {{\partial }_{t}}\Phi  \right|}^{2}}}{\left( \frac{2\rho \ell}{{{\delta }_{\text{eff}}}\w } \right)} \\ 
 & =\frac{C}{8}\frac{
 \mu_{0}^{2}\delta^{3}\ell \w}{\rho}{{\left| {{\left( {{\partial }_{t}}\vec{m} \right)}_{x}} \right|}^{2}},  
\end{align}	
with ${{\delta }_{\text{eff}}}:={C\delta }/{2}$, where $0\le C\le 1$ is a phenomenological parameter that accounts for details of the non-uniform eddy-current distribution in the ferromagnet.
Analogous to the derivation of the radiative damping, we now need the energy of the magnetic excitations. The number of magnons in the system is given by

\begin{equation}
{{N}_{\text{mag}}}=\frac{{{M}_{s}}}{g{{\mu }_{B}}{{\omega }^{2}}}{{\left| {{\partial }_{t}}\vec{m} \right|}^{2}}.
\end{equation}

Thus, the total magnon energy is

\begin{align}
  & {{E}_{\text{mag}}}=\hbar \omega {{N}_{\text{mag}}}\w\ell {\delta} \\ 
 & =\frac{{{M}_{s}}}{\gamma \omega }{{\left| {{\partial }_{t}}\vec{m} \right|}^{2}}\w\ell \delta .  
\end{align}	

The rate of energy dissipation is then given by

	\begin{equation}	\frac{1}{{{T}_{1}}}=\frac{{{P}_{\text{ind}}}}{{{E}_{\text{mag}}}}=\frac{C}{8}\frac{\gamma \omega \mu _{0}^{2}{{M}_{s}}{{\delta }^{2}}}{\rho }\frac{{{\left| {{\left( {{\partial }_{t}}\vec{m} \right)}_{x}} \right|}^{2}}}{{{\left| {{\partial }_{t}}\vec{m} \right|}^{2}}}.
\end{equation}	

The maximum energy decay rate occurs when ${{\left( {{\partial }_{t}}\hat{m} \right)}_{x}}=\left| {{\partial }_{t}}\hat{m} \right|$, in which case

\begin{equation}
{{\left( \frac{1}{{{T}_{1}}} \right)}^{x}}=\frac{C}{8}\frac{\gamma \omega \mu _{0}^{2}{{M}_{s}}{{\delta }^{2}}}{\rho },
\end{equation} 	

where the superscript indicates that this is the maximum decay rate for magnetization oscillations along the $x$-axis. For the case of a perpendicular applied field sufficient to saturate the static magnetization out of the film plane, the damping process is isotropic, i.e.,

\begin{equation}
{{\left( \frac{1}{{{T}_{1}}} \right)}^{y}}=\frac{C}{8}\frac{\gamma \omega \mu _{0}^{2}{{M}_{s}}{{\delta }^{2}}}{\rho }.
\end{equation}	

Therefore, analogous to Eq.~\eqref{deltaomega}, the frequency-swept linewidth $\Delta \omega $ is simply

\begin{align}
  & \Delta \omega =\frac{{{\left( \frac{1}{{{T}_{1}}} \right)}^{x}}+{{\left( \frac{1}{{{T}_{1}}} \right)}^{y}}}{2} \\ 
 & =2{{\alpha }^{\text{eddy}}}\omega,  
\end{align}	

then

\begin{equation}
{{\alpha }^{\text{eddy}}}=\frac{C}{16}\frac{\gamma \mu _{0}^{2}{{M}_{s}}{{\delta }^{2}}}{\rho }.
\end{equation} 	

\bibliographystyle{phd}
\bibliography{references}

\end{document}